\documentstyle[aps,floats,preprint,epsfig]{revtex}

\begin{document}
\draft
\preprint{\vbox{%\hbox{OCIP/C 00-01}
%\hbox{LMU 14/99}
\hbox{hep-ph/0008157}
\hbox{Aug 2000}}}
\title{Discovery and Identification of 
$W'$ Bosons in $e \gamma \to \nu q +X$}
\author{Stephen Godfrey, Pat Kalyniak and Basim Kamal}
\address{
Ottawa-Carleton Institute for Physics \\
Department of Physics, Carleton University, Ottawa, Canada K1S 5B6}
 \author{M.A. Doncheski}
\address{Department of Physics, Pennsylvania State University, \\
Mont Alto, PA 17237 USA}
 \author{Arnd Leike}
\address{Ludwigs--Maximilians-Universit\"at, Sektion Physik, 
Theresienstr. 37,\\
D-80333 M\"unchen, Germany}

\maketitle

\begin{abstract}
We examine the sensitivity of the process 
$e \gamma \rightarrow \nu q +X $ 
to $W'$ bosons which arise in various extensions of the standard model.  
We consider  photon spectra from both the
Weizs$\ddot{a}$cker Williams process and from a backscattered laser.  
The process is found to be sensitive to $W'$ masses up to several TeV, 
depending on the model, the center of mass energy, the 
integrated luminosity, and assumptions regarding systematic errors.  If
extra gauge bosons were discovered 
first in other experiments,
the process could also be used to measure 
$W'$ couplings. This measurement would provide
information that could be used to unravel the underlying theory,
complementary to measurements at the Large Hadron Collider.
\end{abstract}

\section{Introduction}

Extra  gauge bosons, both charged ($W'$) and/or neutral ($Z'$),
arise in many models of physics beyond the Standard 
Model (SM) \cite{c-g,al,sg}. 
Examples include extended gauge theories such as grand unified 
theories \cite{guts} and Left-Right symmetric models \cite{lrmodels} 
along with the corresponding supersymmetric models, 
and other models such as those with finite size extra dimensions
\cite{led}.
To elucidate what physics lies beyond the Standard Model it is necessary 
to search for manifestations of that new physics
with respect to the predicted particle 
content.  Such searches are
a feature of ongoing collider experiments and the 
focus of future experiments.  The discovery of  new particles  would 
provide definitive evidence for physics beyond the Standard Model and, 
in particular, the discovery of new gauge bosons would indicate that 
the standard model gauge group was in need of extension.
There is a considerable literature on $Z'$ searches.  In 
this paper we concentrate on $W'$ searches, for which less work has 
been done.

Limits have been placed on the existence of new gauge bosons through 
indirect searches based on the
deviations from the SM they would produce in precision electroweak 
measurements. 
For instance, indirect limits from $\mu$-decay constrain the
left-right model $W'$ 
to $M_{W_{R}} \gtrsim 550$ GeV \cite{Barenboim}.  A more severe
constraint 
arises from $K_L - K_S$ mass-splitting which gives $M_{W_{R}} \gtrsim 
1.6$~TeV \cite{pdb}, assuming equal
coupling constants for the two $SU(2)$ gauge groups.

New gauge boson searches at hadron colliders consider their 
direct production via the Drell-Yan process and their subsequent decay 
to lepton pairs. For $W'$ bosons, decays to hadronic jets are 
sometimes also considered.  The 
present bounds on $W'$ bosons  from the CDF and D0 
collaborations at the Tevatron $p\bar{p}$ collider at Fermilab are 
$M_{W'} \gtrsim 720$~GeV \cite{pdb}.  
The search reach is expected to increase by $\sim 
300$~GeV with 1~fb$^{-1}$ of luminosity \cite{tev-lhc}.  
The Large Hadron Collider is 
expected to be able to discover
$W'$'s up to masses of $\sim 5.9$~TeV \cite{tev-lhc}.  
These $W'$ limits assume SM 
strength couplings and decay into a light stable neutrino 
which is registered in the detector as missing $E_T$.  They can 
be seriously degraded by loosening the model assumptions.

There are 
few studies of indirect searches for $W'$ bosons at $e^+e^-$
colliders. In a recent paper we examined the sensitivity of the
reaction $e^+e^- \to \nu \bar{\nu} \gamma$ to 
$W'$ and $Z'$ bosons\cite{wp1} (see also \cite{hewett}).  
We found it was sensitive to $W'$'s up 
to several TeV in mass, depending on the model, the centre of mass 
energy, and the integrated luminosity.

Clearly, if deviations from the standard model are observed, it will 
take many different measurements to disentangle the nature of the new 
physics responsible.  In this paper we present the results of a study 
of the sensitivity of the process $e \gamma \to \nu q +X$ to $W'$ 
bosons.  We find that this process is sensitive to 
$W'$ masses up to several TeV,
in many cases more sensitive than the process $e^+e^- \to \nu \bar{\nu}
\gamma$.
In particular,  we find that it is far more sensitive to the 
Un-Unified model (UUM)  \cite{uum}. 
% although, on the other hand, it is less 
% sensitive to the Left-Right Model (LRM).
The process $e \gamma 
\to \nu q +X$ is sensitive to both the quark and lepton couplings to 
the $W'$ and, furthermore, does not have the complication of contributions
from $Z'$ 
bosons.  It can therefore contribute information that complements that 
obtained from $e^+e^- \to \nu \bar{\nu} \gamma$, helping to build a 
picture of the underlying physics.

We  consider various extended 
electroweak models.  The first is the 
Left-Right symmetric model \cite{lrmodels}
based on the gauge group 
$SU(3)_C \times SU(2)_L \times SU(2)_R \times U(1)_{B-L}$ which has 
right-handed charged currents. 
The second is the Un-Unified model 
\cite{uum} which is based on the gauge group $SU(2)_q \times SU(2)_l 
\times U(1)_Y$ where the quarks and leptons each transform under their own 
$SU(2)$.  The final type of model, denoted as the KK model, 
contains the Kaluza-Klein excitations of the SM gauge bosons which 
are a possible consequence of theories with large extra dimensions
\cite{led}.  
We give the relevant couplings for those models in
Section II but refer the reader to the literature for more details.
Additionally, we study discovery limits for a $W'$ boson with SM 
couplings.  Although it is not a realistic model, this so-called
Sequential Standard Model (SSM) has been
adopted as a benchmark 
to compare the discovery reach of different processes.

We will find that the process $e \gamma 
\to \nu q +X$ can 
indeed extend the discovery reach for $W'$'s significantly beyond 
$\sqrt{s}$, with the exact limit depending on the specific model. 
Additionally,
if extra gauge bosons are discovered which are not overly 
massive, the process considered here could be used to
measure their couplings. This 
would be crucial for determining the origins of the  $W'$.  As 
such, it would play an important complementary role to the LHC studies.

In the next section, we describe the details of our
calculations. The resulting $W'$ discovery limits and projected
sensitivities for 
$W'$ couplings are given in Section III.  We conclude with some 
final comments.

\section{Calculations and Results}

\begin{figure}
\centerline{\epsfig{file=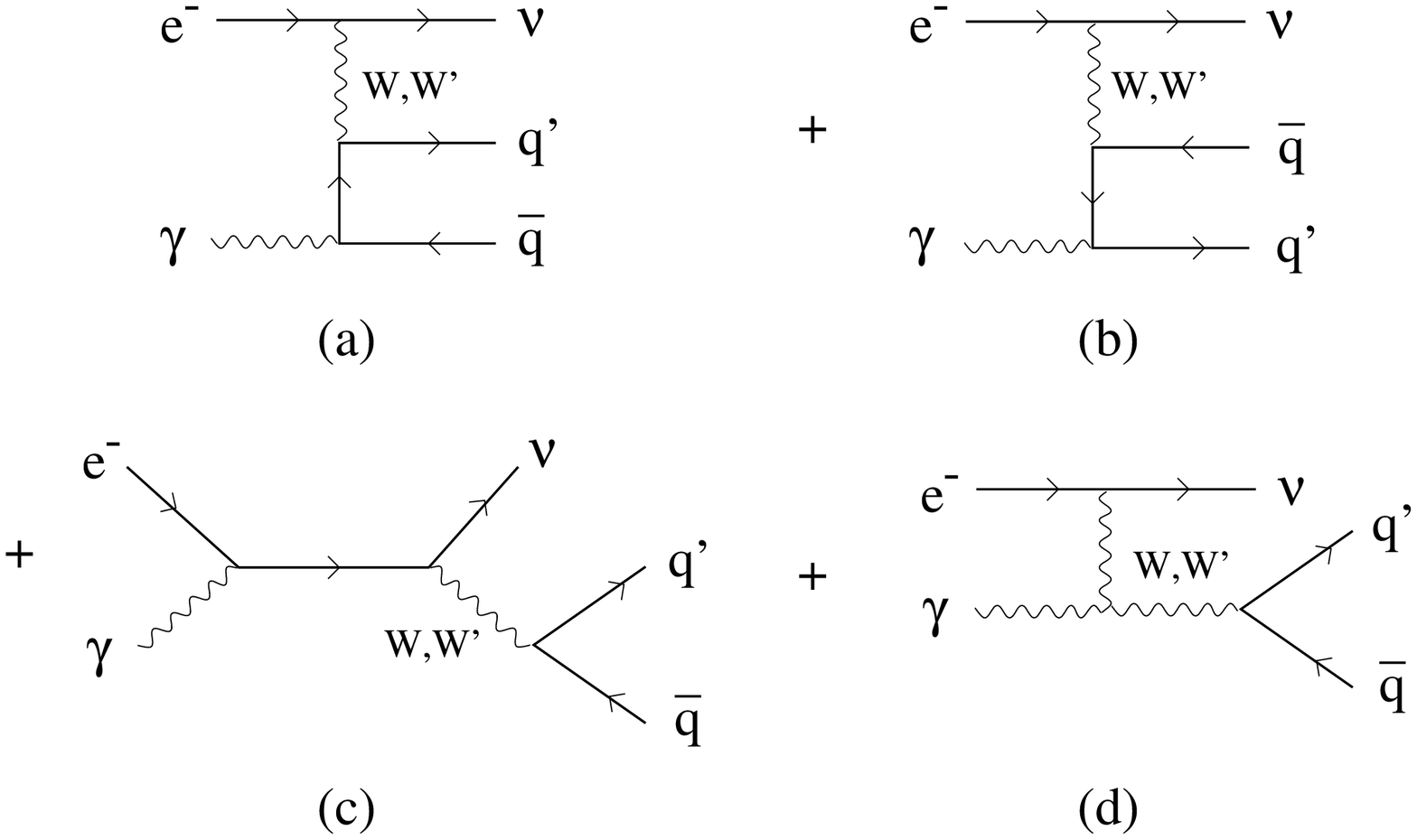,width=6.5in}}
\vspace{20pt}
\caption{The Feynman diagrams contributing to the process
$e \gamma \to \nu q\bar{q}$.}
\label{Fig1}
\end{figure}

\begin{figure}
\centerline{\epsfig{file=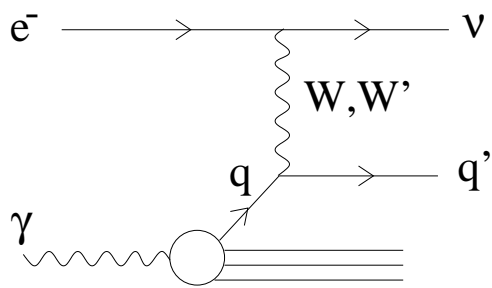,width=2.5in}}
\vspace{20pt}
\caption{The Feynman diagram contributing to the process
$e q \to \nu q'$.}
\label{Fig2}
\end{figure}

We are interested in the process 
\begin{equation}
\label{process1}
e\gamma \to \nu q \bar{q}
\end{equation}
where the photon arises from either a backscattered laser 
\cite{backlaser}
or from 
Weizs$\ddot{a}c$ker Williams bremsstrahlung \cite{WW}
from the incident $e^{+(-)}$
beam.
The relevant Feynman diagrams are given in Fig.~\ref{Fig1}.
In this process it is diagrams 1(a) and 1(b) with the $W'$ exchanged in
the t-channel which are most sensitive to the effects of the $W'$.  The 
contribution of these diagrams can be enhanced by imposing the kinematic cut 
that one of either the $q$ or the $\bar{q}$ is collinear to the beam axis.
In this kinematic 
region the process $e\gamma \to \nu q \bar{q}$ is approximated quite 
well by the simpler process
\begin{equation}
\label{process2}
e q \to \nu q'
\end{equation}
shown in Fig.~\ref{Fig2}, 
where the quark is described by the quark parton content of the 
photon, the so-called resolved photon approximation 
\cite{resphot,grv2,SaS}.
This has been
verified numerically by comparing kinematic 
distributions of the outgoing quark calculated using both process 
\ref{process1} and process \ref{process2} 
for a given detector acceptance where the outgoing $q$ ($\bar{q}$) is
constrained 
to $|\cos\theta_{q(\bar{q})}| \leq \cos \theta_{cut}$ and in addition 
for process \ref{process1}  $|\cos\theta_{\bar{q}(q)}| \geq \cos 
\theta_{cut}$ (where $\theta_{cut} \simeq 10^0$),
i.e.\ in process 1, one jet is observed while the other is lost down the beam 
pipe.
We use the process $e q \to \nu q'$ to obtain limits as it is 
computationally much faster and the discovery limits obtained 
using this approximation are in good agreement with those using the 
full process.  The reliability of the limits have been further checked 
by using different photon distribution functions.

The expression for the unpolarized cross section is given by
\begin{equation}
\sigma= \int dx \int dy f_{\gamma/e}(x,\sqrt{s}/2) f_{q/\gamma}(y,Q^2) 
\hat{\sigma}(e q \to \nu q')
\end{equation}
where $f_{\gamma/e}(x)$ is the photon distribution,
$f_{q/\gamma}(y) $ the distribution for the quark content in the 
photon, and $\hat{\sigma}(e q \to \nu q')$ is the cross section for 
the parton level process given by:
\begin{equation}
\hat{\sigma}(e^- q \to \nu q')=\int d\hat{t} {{d\hat{\sigma}}\over 
{d\hat{t}}}
\end{equation}
where 
\begin{equation}
\label{dsdt}
{{d\hat{\sigma}}\over {d\hat{t}}}={{\pi \alpha^2}\over{4 \sin^4\theta_w}}
\times f(\hat{s},\hat{u})
\end{equation}
and
\begin{eqnarray}
\label{fsu}
f(\hat{s},\hat{u}) & = & {1\over{(\hat{t}-M_W^2)^2}  }
\left\{ { 1+2 C_L^q C_L^l \left( { {\hat{t}-M_W^2}\over{\hat{t}-M_{W'}^2}
} \right)
} \right. \nonumber \\
& & +\frac{1}{2} \left( { {\hat{t}-M_W^2}\over{\hat{t}-M_{W'}^2} }
\right)^2
\left[ { ({C_L^q}^2 + {C_R^q}^2 )   ({C_L^l}^2 + {C_R^l}^2
)(1+\hat{u}^2/\hat{s}^2)
} \right.
\nonumber \\
& & \qquad 
\left. { \left. {
+({C_L^q}^2 - {C_R^q}^2 )   ({C_L^l}^2 - {C_R^l}^2
)(1-\hat{u}^2/\hat{s}^2)
}\right] } \right\}
\end{eqnarray}
and $\hat{s}$, $\hat{t}$, and $\hat{u}$ are the usual Mandelstam variables
for the parton level process. We take $Q=\sqrt{s_{eq}}$ in $f_{q/\gamma}$ and
the scale $\sqrt{s}/2$ in $f_{\gamma/e}$ is only relevant for photons produced
via the Weizs$\ddot{a}$cker Williams process.
The process 
$e^- \bar{q} \to \nu \bar{q}'$ also contributes to the same 
experimental signature.  Its cross section is given by Eq.
\ref{dsdt} but with $\hat{s}$ and $\hat{u}$ interchanged in Eq.~\ref{fsu}
such
that 
$f(\hat{s},\hat{u}) \Leftrightarrow f(\hat{u},\hat{s})$.  
Similarly,  for the processes $e^+ \bar{q} \to \bar{\nu} \bar{q}'$ 
and $e^+ q \to \bar{\nu} q'$, which contribute in the case the of the
Weizs$\ddot{a}$cker Williams process, the cross section 
is given with $f(\hat{s},\hat{u})$ and $f(\hat{u},\hat{s})$ 
in Eq.~\ref{fsu} respectively.
Our conventions for 
the couplings, 
$C_L$ and $C_R$, follow from the vertices
\begin{equation}
\label{Wcoup}
W_i f \bar{f}' = \frac{ig}{\sqrt{2}} \gamma^\mu \left(
\frac{1-\gamma_5}{2}
\,\,C_L^{W_i} + \frac{1+\gamma_5}{2} \,\,C_R^{W_i} \right) .
\end{equation}
Thus, in the SM,  $C_L^{W_1}=1$, and $C_R^{W_1}=0$. A $W'$ in the SSM also
has these SM couplings. In the case of the KK model, the couplings are  
enhanced by a factor of $\sqrt{2}$ such that $C_L^{W_{KK}}=\sqrt{2}$, and
$C_R^{W_{KK}}=0$. In the LRM, the extra $W_R$ has only right-handed
couplings such that $C_L^{W_R}=0$, and $C_R^{W_R}=\kappa$. Here the
parameter $\kappa = g_R/g_L$ is the  ratio of the coupling constants of
the two $SU(2)$ gauge groups. Since we will ultimately find that the
process under consideration here is not as sensitive to a $W_R$ as some
other processes, 
we will only consider the LR model for
$\kappa = 1$. We also take the CKM matrix elements 
for right-handed fermions to be equal to those of left-handed fermions.
In each of the models mentioned
so far, the couplings of the $W'$ to the quarks and the leptons are equal.
In the case of the UUM, we have instead
$C_L^{l}=-\frac{\sin\phi}{\cos\phi}$,
$C_L^{q}=\frac{\cos\phi}{\sin\phi}$, and $C_R^{l}=C_R^{q}=0$. 
The UUM is parametrized by an angle $\phi$, which
represents
the
mixing of the charged gauge bosons of the two $SU(2)$ groups. The process
we consider is actually insensitive to the parameter $\phi$ because
$C_L^{l}$ and $C_L^{q}$ always multiply each other 
in the expressions for the cross section.
The polarized 
cross sections may be inferred from the coupling structure in 
Eq.\ (\ref{fsu}).

We begin by showing and discussing 
the total cross sections and the differential cross
sections $d\sigma/dE_q$ and $d\sigma/{dp_T}_q$.  We do not show the 
angular distribution as it gives lower limits than do the $E_q$ and 
${p_T}_q$ distributions.   We take the SM inputs
$M_W = 80.33$ GeV, $\sin^2\theta_W = 0.23124$, and $\alpha=1/128$ 
\cite{pdb}.
Since we work only to leading order,
there is some arbitrariness in the above input,
in particular $\sin^2\theta_W$.  

To take into account detector acceptance, the angle of the observed jet,
$\theta_{q(\bar{q})}$, has been restricted to the range
\begin{equation}
\label{accept}
10^0 \leq \theta_{q(\bar{q})} \leq 170^0.
\end{equation}
In extracting limits we 
will also restrict the jet's transverse momentum to reduce hadronic 
backgrounds which we discuss below.

\begin{figure}
\centerline{\epsfig{file=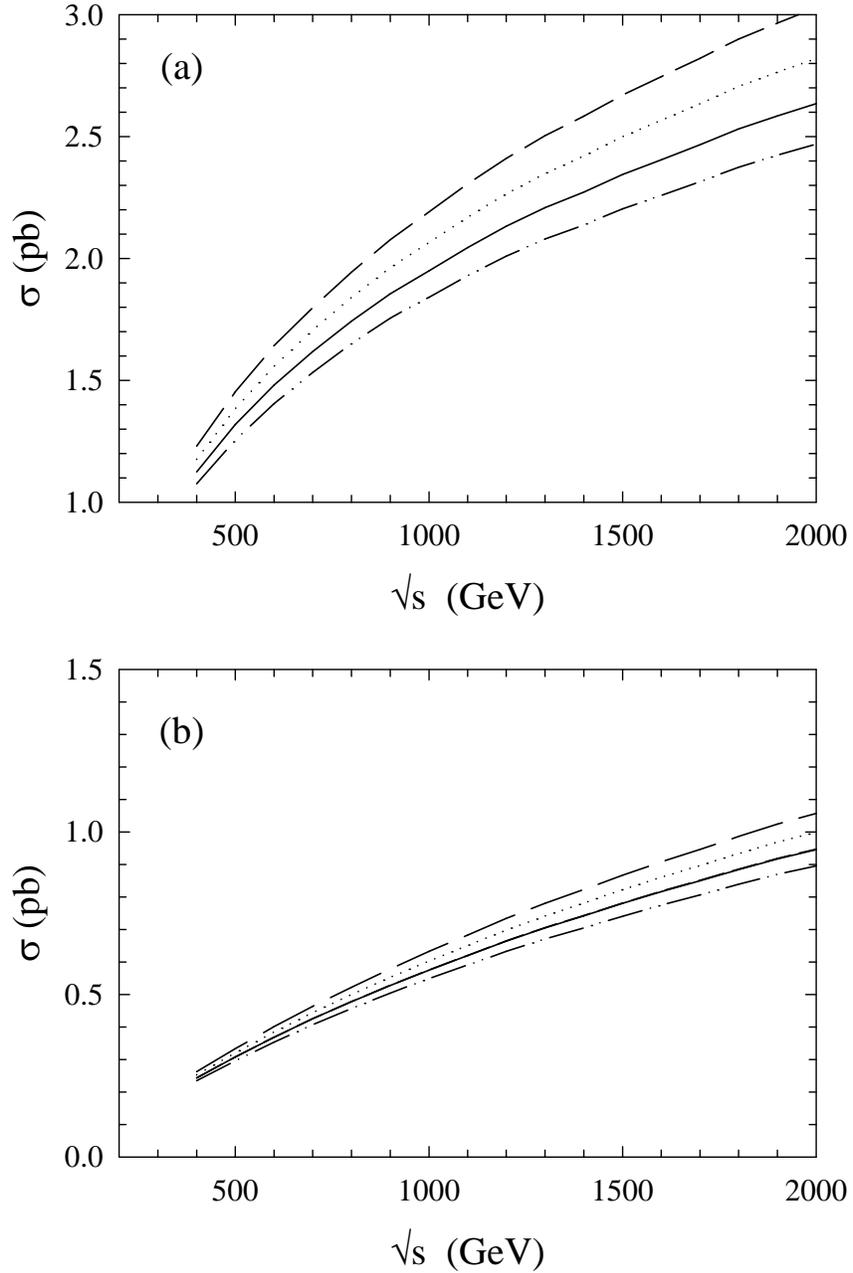,width=4.5in}}
\vspace{20pt}
\caption{The cross section 
$\sigma(e^-\gamma \to \nu q +X) $
as a function of $\sqrt{s}$ for the SM 
(solid line),  and with a $W'$ of mass 750~GeV for the SSM 
(dotted line), UUM (dash-dot-dot line), and the KK model (dashed line).  
The LR model cross section overlaps the SM cross section.
(a) uses the backscattered laser photon spectrum and 
(b) uses the Weizs\"acker-Williams photon spectrum.}
\label{Fig3}
\end{figure}

The unpolarized cross sections, $\sigma$, for photons coming from the 
backscattered laser photon 
distribution and the Weizs$\ddot{a}$cker Williams distributions
are shown in Fig.~\ref{Fig3}.  We have included $u$, $d$, $s$ and
$c$-quark 
contributions and used the leading order GRV distributions in calculating 
these cross sections \cite{grv2}.  
We will discuss the use of other 
parametrizations below.  For the case of the backscattered photon we 
included the subprocesses $e^- q \to \nu q'$ and $ e^- \bar{q} \to \nu 
\bar{q}'$ where the $q$ could be either $u$ or $c$ and the $\bar{q}$ 
could be a $\bar{d}$ or $\bar{s}$.  For the Weizs$\ddot{a}$cker Williams
case 
the photon can be radiated from either the $e^-$ or $e^+$ so we must 
also include the subprocesses $e^+ q \to \bar{\nu} q'$ and 
$ e^+ \bar{q} \to \bar{\nu} \bar{q}'$. 
The cross section is shown for the SM, LRM ($\kappa=1$), UUM,
SSM, and KK model, with $M_{W'}=750$ GeV in each case.
The mass choice is rather arbitrary, made to
illustrate general behaviour.  We do not show cross sections for 
polarized electrons.  The cross section for right handed electrons 
($\sigma_R = \sigma(e^-_R)$) only couples to $W'$'s in the LR model.  
In all other cases considered here, $\sigma_R$ is zero.

One first notes that the cross sections for the backscattered laser 
case are somewhat larger than those for the $e^+e^-$ case with WW photons.
This is a
direct 
consequence of the harder photon spectrum in the case of the former.
The cross sections are typically of the order of several picobarns.  
For the luminosities expected at high energy $e^+e^-$ 
colliders this results in statistical errors of less than a percent.  
With systematic errors expected to be of the order of 2\% we therefore
expect systematic errors to dominate over statistical errors.  From 
Fig.~\ref{Fig3} one also sees that, at least for the example shown,
the deviations due to a $W'$ are significantly larger than the 
expected measurement error.  Thus, it appears that this process will 
provide a sensitive probe for $W'$ bosons.

\begin{figure}
\centerline{\epsfig{file=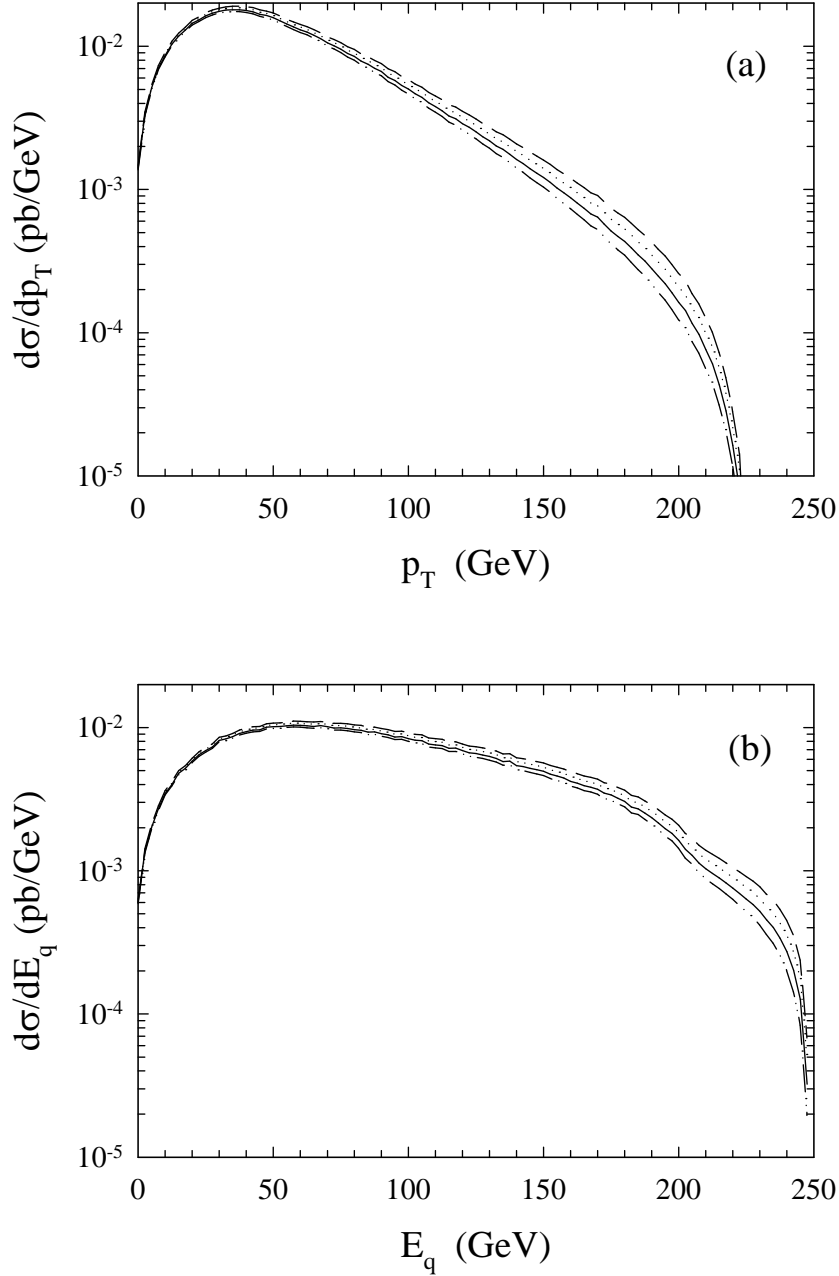,width=4.5in}}
\vspace{20pt}
\caption{The differential cross sections   (a) $d\sigma/d{p_T}_q$ and 
(b) $d\sigma/dE_q$.  They are shown for the backscattered laser case
with $\sqrt{s}=500$~GeV and for $M_{W'}=750$~GeV.  
The SM is given by solid line,   the SSM by the dotted line, 
UUM by the dash-dot-dot line, and the KK model by the dashed line.  
}
\label{Fig4}
\end{figure}

We are interested in reactions in which only the quark (or antiquark) 
jet is observed.
The  kinematic observables of interest are therefore, the jet's energy, 
$E_q$, its momentum perpendicular to the beam axis, ${p_T}_q$, and its 
angle relative to the incident electron, $\theta_q$, all defined in 
the $e^+e^-$ center-of-mass frame.  The differential cross sections,
$d\sigma/{dp_T}_q$ and $d\sigma/dE_q$, are shown in Fig.~\ref{Fig4}
for the standard model and  the SSM, LRM, UUM, and KK models 
for the backscattered laser case with
$\sqrt{s} = 500$ GeV and $M_{W'}=750$~GeV.  We do not show the 
angular distribution as we found that the
$d\sigma/{dp_T}_q$ and $d\sigma/dE_q$ distributions were more 
sensitive to $W'$'s.  We see that extra $W$ bosons
result in larger {\sl relative} deviations from the SM
in the high $E_q$ or ${p_T}_q$ regions but since the lower 
$E_q$ or ${p_T}_q$ regions have higher statistics the result is
roughly similar in significance in both kinematic regions although the high
${p_T}_q$ ($E_q$) region will be less affected by systematic error.
To maximize 
the potential information we divide the distributions into 10 
equal sized bins and calculate the $\chi^2$ by summing over the bins.  

\begin{figure}
\centerline{\epsfig{file=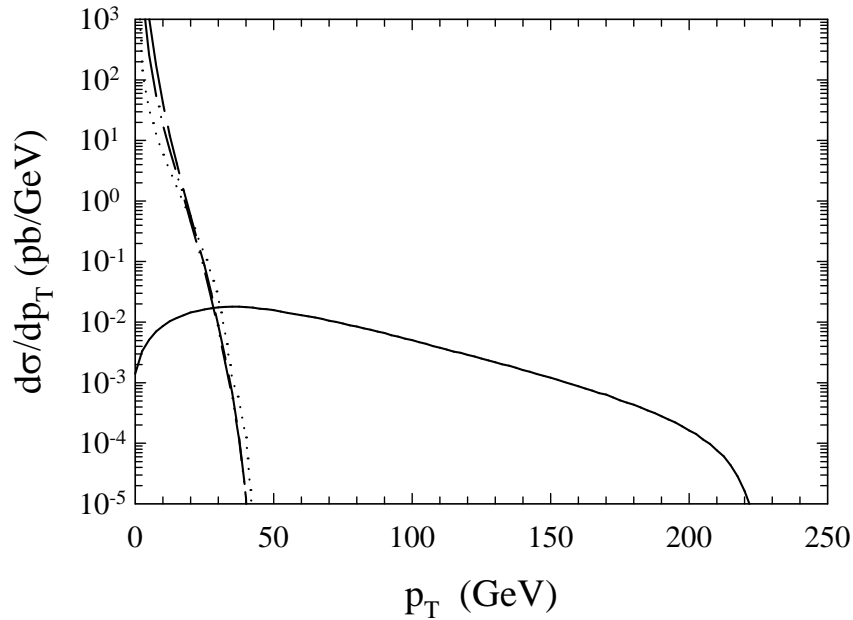,width=4.5in}}
\vspace{20pt}
\caption{The differential cross section   $d\sigma/d{p_T}_q$ for the 
SM and the various backgrounds.  
They are shown for the backscattered laser case
with $\sqrt{s}=500$~GeV.  The process $e\gamma\to \nu q +X$ is 
given by the solid line,   the subprocess $\gamma\gamma \to q\bar{q}$
by the dotted line, the singly resolved backgrounds by 
the dash-dot-dot line, and the doubly resolved backgrounds
by the dashed line.  For the backgrounds we impose the cuts that one 
jet is observed with $|\cos\theta_q| < \cos 10^o$ while the other jet 
is lost down the beampipe with $|\cos\theta_q| > \cos 10^o$.
}
\label{Fig5}
\end{figure}

Before proceeding to our results
we must deal with the issue of backgrounds.  The dominant backgrounds
arise from two jet final states where one of the jets goes down the 
beam pipe and is not observed.  Processes which contribute two jet final 
states are:  $\gamma\gamma \to q\bar{q}$, the once resolved reactions
$\gamma g \to q \bar{q}$ and $\gamma q \to g q$, and the twice 
resolved reactions $g g \to q\bar{q}$, $q\bar{q}\to q\bar{q}$, $qg 
\to q g$ ... \cite{bcontent}.  
In addition there are backgrounds involving t-channel 
exchange of massive gauge bosons but these are suppressed relative to 
the backgrounds already listed.   In Fig.~\ref{Fig5} we show the ${p_T}_q$ 
distributions for these backgrounds 
with only one of the jets observed and 
the other going down the beam pipe for $\sqrt{s}=500$~GeV for the 
backscattered laser case.
We use the criteria that, to be 
seen, the parton must satisfy $170^0 \geq \theta_{q(\bar{q})} \geq 10^0$.  
It is likely that these cuts would be more stringent in a real 
detector and with veto detectors close to the beam pipe. However, other 
issues such as spread of the hadronic jets and the remnants of the 
photon complicate the analysis and must be carefully considered.  
In the absence of a detailed detector simulation we feel that the chosen 
detector cuts will give a reasonable representation of the situation 
for the purposes of estimating the discovery potential of this 
process.  To extract limits from real data these effects must, of 
course, be studied in detail. 
Referring to Fig.~\ref{Fig5}, the constraint that ${p_T}_q \geq 40$~GeV 
effectively eliminates these backgrounds.  Similarly,  for 
$\sqrt{s}=1$~TeV we take ${p_T}_q \geq 75$~GeV and for 
$\sqrt{s}=1.5$~TeV we take ${p_T}_q \geq 100$~GeV.  

\subsection{Discovery Limits for $W'$'s}

The best discovery limits were in general obtained using the observable
$d\sigma/d{p_T}_q$ with $d\sigma/d{E}_q$ being only slightly less 
sensitive.  We found that limits obtained using other observables 
such as the total cross section, forward backward asymmetry and
$d\sigma/d\cos\theta_q$ were less sensitive 
so the results we present will be based on the ${p_T}_q$ 
distributions.  In addition, for the backscattered laser case
limits were obtained for the LR model 
using the right-handed polarized cross section, $\sigma_R$ and the
left-right asymmetry,
\begin{equation}
\label{alr}
A_{LR} = \frac{\sigma_L-\sigma_R}{\sigma_L+\sigma_R}.
\end{equation}
For 100\% polarization and only including statistical errors one 
obtains reasonable limits for the LRM $W'$.  However, for more 
realistic polarizations of 90\% these limits are seriously degraded and 
only slightly greater than those obtained from unpolarized 
${p_T}_q$ and $E_q$ distributions.  Once systematic errors are 
included,  even though they were only one 
half those used in the $d\sigma/d{p_T}_q$ calculation (since one expects 
some cancellation of errors between the numerator and denominator in 
$A_{LR}$), we find that the sensitivity of $A_{LR}$
is less than those obtained 
from the ${p_T}_q$ and $E_q$ distributions while the sensitivity of $\sigma_R$
is roughly comparable to the sensitivity of the distributions.  
We therefore only 
report  limits obtained  from the distributions  in Tables I and II.

In obtaining the $\chi^2$ for $d\sigma/d{p_T}_q$, we used 10 equal
sized  bins in the range ${p_T}_q^{\rm min} < {p_T}_q <
{p_T}_q^{\rm max}$, where ${p_T}_q^{\rm min}$ is given by the
${p_T}_q$ cut chosen to reduce the two jet backgrounds and
${p_T}_q^{\rm max}$ is taken to be the kinematic limit. 
We have
\begin{equation}
\label{sumbin}
\chi^2 = \sum_{\rm bins} \left(\frac{d\sigma/d{p_T}_q - 
d\sigma/d{p_T}_{q, {\rm SM}}}{\delta d\sigma/d{p_T}_q}\right)^2,
\end{equation}
where $\delta d\sigma/d{p_T}_q$ is the error on the measurement.
Analogous formulae hold for other observables. One sided 95\% confidence
level discovery limits are obtained by requiring $\chi^2\geq 2.69$
for discovery. Systematic errors, when included,
were added in quadrature with the statistical errors.

The discovery limits for all five models are listed in
Table~\ref{limitstabbl} for the
backscattered laser case and in Table~\ref{limitstabww} for the $e^+e^-$
case with WW
photons. Results are presented for $\sqrt{s}=0.5$, 1.0 and 1.5 TeV, using
the same input
parameters as for the cross sections presented in the previous section. 
For each center-of-mass energy, two luminosity scenarios are considered
and we present
limits obtained with and without systematic errors. Our prescription is
to include a 2\% systematic error per bin. This number is quite
arbitrary but seems reasonable.
In addition to detector systematics, which we
expect will dominate, there are uncertainties associated with the
beam luminosity and energy, which will be spread over a range. Other
systematic errors are associated with background subtraction 
as well as radiative corrections.
Thus,  the 2\% number should not be taken too seriously 
except to highlight the fact that a precision measurement is
required to take full advantage of the large event rate.

The discovery limits are substantial and compare favourably
in most cases
 to those 
obtained from the previously studied process $e^+e^- \to \nu 
\bar{\nu}\gamma$.  
In all cases, the limits for the WW process are significantly lower than
with the backscattered laser. This enhanced reach is an argument in favour
of $e \gamma$ colliders. As expected, when systematic errors are not included 
there is a significant improvement in the limits with the higher 
luminosity.  When 2\% systematic errors are included the improvement 
for the high luminosity scenario is not as dramatic.  

In every case for the process considered here, the SSM and UUM yield the
same discovery limits. This occurs because the two processes represent
positive and negative interferences of equal strength, respectively, with
the SM. The limits obtained here for the
backscattered laser case for the SSM are similar to those from the 
process $e^+e^- \to \nu 
\bar{\nu}\gamma$ before a systematic error is included. However, the
limits here are downgraded less than for the previous process when a 2\%
systematic error is included, resulting in a higher discovery reach for
this process. The behaviour is similar for the LRM. The limits, for the
backscattered laser case, including systematic errors, are similar to
those obtained for $e^+e^- \to \nu \bar{\nu}\gamma$.

For both the KK model and the UUM, this process offers a significant
improvement over the process $e^+e^-\to \nu\bar{\nu}\gamma$. 
For the backscattered laser case, the KK $W'$ limits are
typically 2 times higher when a systematic error is included than for the
previous process. Similarly, the limits obtained for the UUM 
with the process
considered here are a factor of 3 better than for the equivalent case in
the previous process.

Beam polarization of 90\% does not improve the limits.  The only model 
studied which would benefit from polarization is the LRM for which the 
$W'$ is right-handed.  The process $e\gamma \to \nu  q$ proceeds 
via t-channel $W$-exchange.  The SM contribution is totally left-handed
so that for 100\% right-handed polarization a $W_R$ would show up 
quite dramatically.  However, the cross-section for 
$W_R$'s which are not unrealistically light is orders of magnitude smaller
than 
the left-handed cross section.  So even a  small pollution  of 
left-handed electrons would largely overwhelm the right-handed cross 
section.  The signal from right-handed $W'$'s is further eroded when 
systematic errors are included.  Quantitatively we found for the back 
scattered laser case with $\sqrt{s}=500$~GeV and $L=50$~fb$^{-1}$ 
using 100\% right-handed electrons a discovery limit of 1.7~TeV.  This 
was degraded to $\sim 700$~GeV when 90\% polarization was used and was 
further degraded to $\sim 600$~GeV when a 2\% systematic error on the 
cross section was included.  This is only slightly higher than
what is obtained using the 
${p_T}_q$ distribution with unpolarized electrons.  

The UUM  is an interesting case.  First, 
as noted above,
the process $e\gamma \to 
\nu  q$ is considerably more sensitive to a UUM $W'$ than the
process studied previously, $e^+e^-\to \nu\bar{\nu}\gamma$.  In 
addition, the dependence on 
the mixing angle between the two $SU(2)$ groups cancels in the process 
we are studying. The cross section exhibits a straightforward destructive
interference of the $W'$ exchange with the SM $W$ exchange, rather than
the complicated $\phi$ dependence of the 
process  $e^+e^-\to \nu\bar{\nu}\gamma$.  This is an example of 
how the two processes complement each other.

As indicated above, we also derived discovery limits using another
parametrization of the photon distribution functions, that of Schuler and
Sj\"{o}strand \cite{SaS}. 
That particular distribution results in lower cross
sections than does the GRV parametrization. However, the discovery limits
were consistently within 50-100~GeV of those reported in the Tables.

\subsection{Constraints on Couplings}

In this section, we consider constraints
on the couplings of an extra $W'$ from the process 
$e \gamma \to \nu q +X$.  All constraints are shown
for the backscattered laser case.
These constraints are significant only in the case where 
the mass of the corresponding extra gauge boson is 
considerably lower than its search limit in this process.
We assume here that a signal for an extra gauge boson has been detected 
by another experiment, such as at the LHC.

Given such a signal, we derive constraints (at 95\% C.L.) on the
couplings of 
extra gauge bosons. We present the constraints in terms of couplings 
$C_L$ and $C_R$ which are normalized as in Eq.~\ref{Wcoup}

The constraints correspond to 
\begin{equation}
\label{chi2}
\chi^2=\left(\frac{O_i(SM)-O_i(SM+W')}{\delta O_i}\right)^2=5.99,
\end{equation}
where $O_i(SM)$ is the prediction for the observable 
$O_i$ in
the SM, $O_i(SM+W')$ is the prediction of the
extension of the SM and $\delta O_i$ is the expected experimental error.
The index $i$ corresponds to different observables such as
$\sigma$, $\sigma_R$, $A_{LR}$, or $d\sigma/d{p_T}_q$ where one sums over
all 
bins as in Eq.~\ref{sumbin} for the latter.

We examined the sensitivity to polarized beams with the assumption 
that for single
beam ($e^-$) polarization, we have, as in the previous section,
equal running in left and right polarization states. We found, as 
before, that polarized beams with realistic polarization, offer
little improvement over the case of unpolarized beams.  

\begin{figure}
\centerline{\epsfig{file=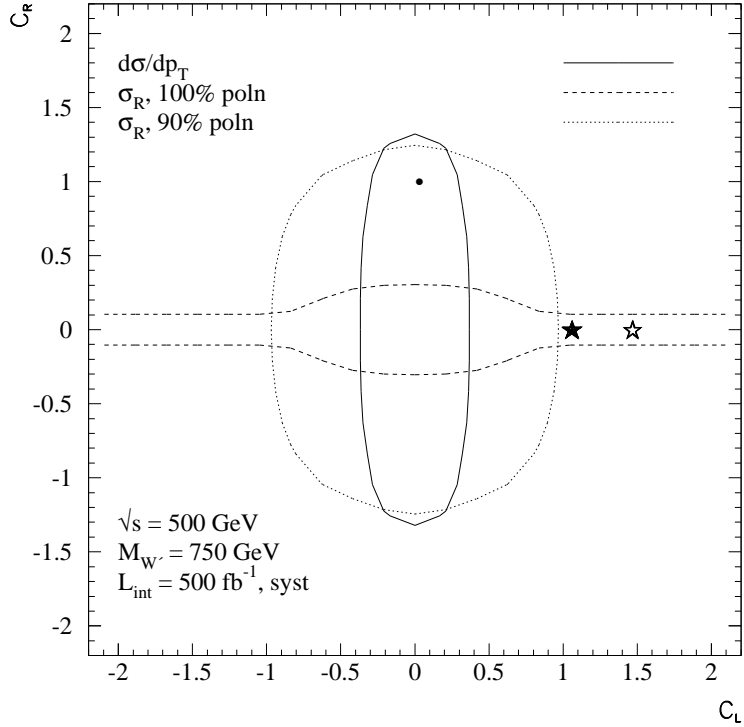,width=4.5in}}
\vspace{10pt}
\caption{95\% C.L. constraints on $W'$ couplings arising from
$d\sigma/dp_T$ (solid line) and $\sigma_R$ with 100\% (dashed line) and
90\% (dotted line) polarization. The results are for $\sqrt{s}=500$~GeV with 
the backscattered laser spectrum,
$M_W'$ = 750~GeV, and $L_{int}=500$~$fb^{-1}$ with a 2\% systematic error. 
The couplings corresponding to the SSM, LRM and the KK model are indicated 
by a full star, a dot and an open star, respectively.
} 
\label{Fig6}
\end{figure}

\begin{figure}
\centerline{\epsfig{file=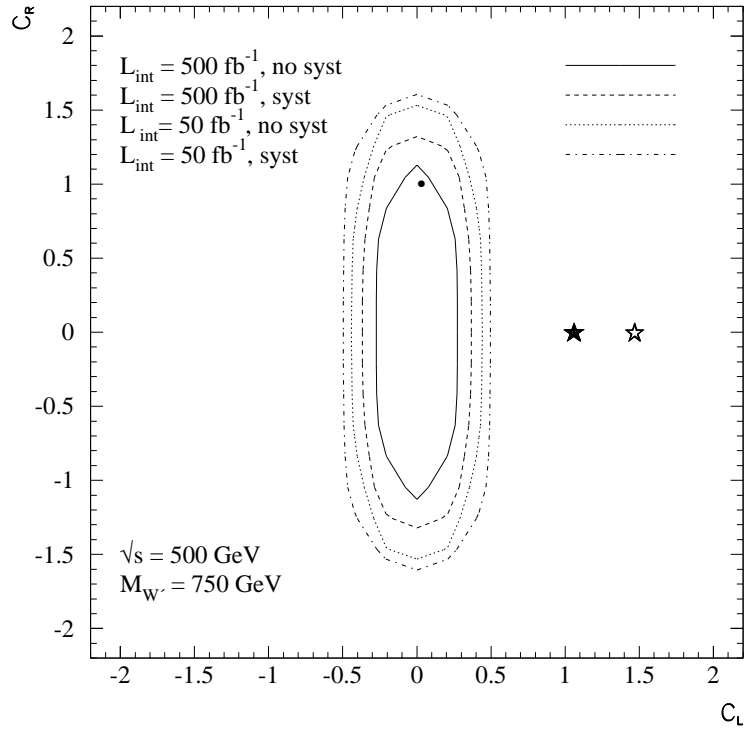,width=4.5in}}
\vspace{10pt}
\caption{95\% C.L. constraints on $W'$ couplings for $\sqrt{s}=500$~GeV
with the backscattered laser spectrum
and
$M_W'$ = 750~GeV. The integrated luminosity and systematic error is
varied.
}
\label{Fig7}
\end{figure}

\begin{figure}
\centerline{\epsfig{file=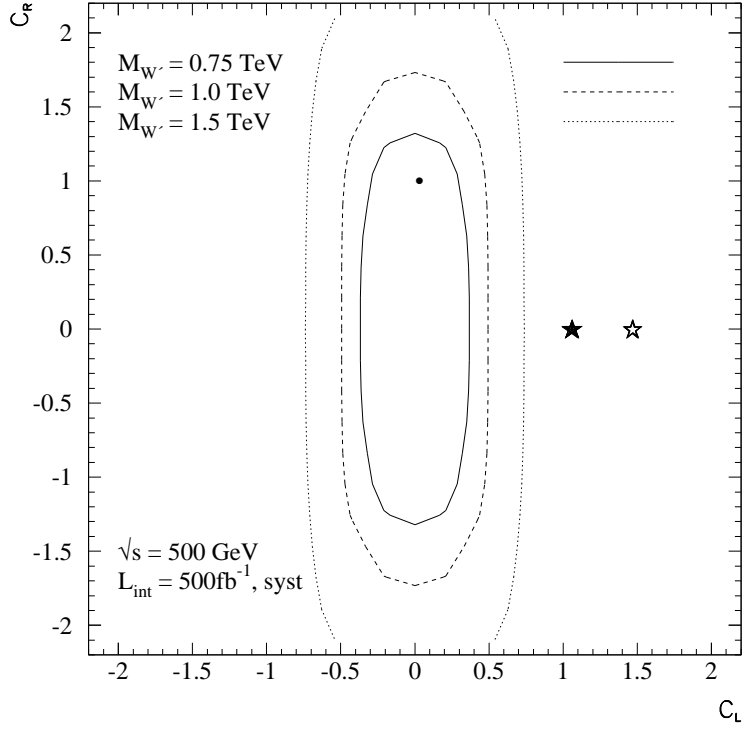,width=4.5in}}
\vspace{10pt}
\caption{95\% C.L. constraints on $W'$ couplings for $\sqrt{s}=500$~GeV
with the backscattered laser spectrum 
and $L_{int}=500$~$fb^{-1}$ with a 2\% systematic error. The three $W'$
masses of 0.75~TeV (solid line), 1.0~TeV (dashed line) and 1.5~TeV (dotted
line) are presented. } 
\label{Fig8}
\end{figure}

\begin{figure}
\centerline{\epsfig{file=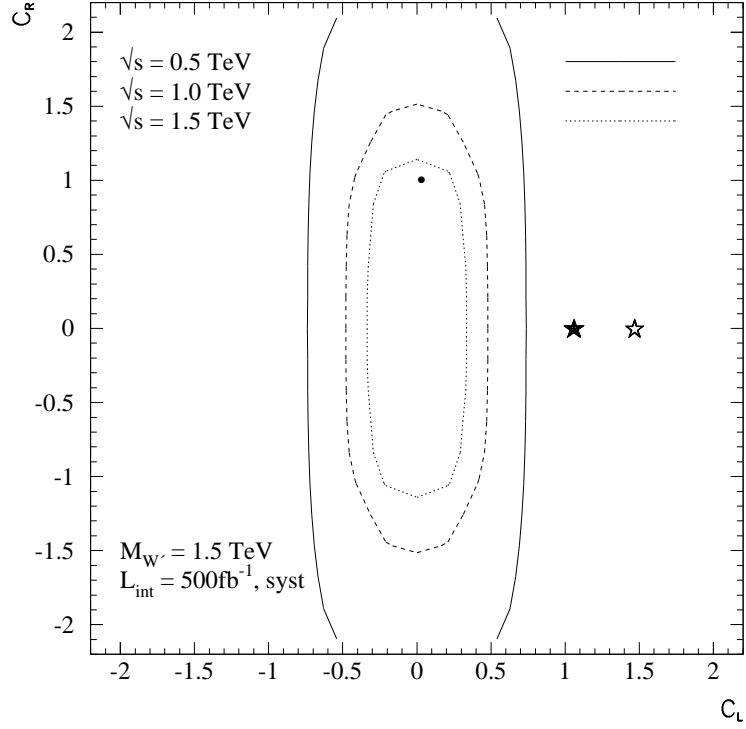,width=4.5in}}
\vspace{10pt}
\caption{95\% C.L. constraints on $W'$ couplings for $M_W'=1.5$~TeV
and $L_{int}=500$~$fb^{-1}$ with a 2\% systematic error. The three
collider energies of 0.5~TeV (solid line), 1.0~TeV (dashed line) and
1.5~TeV (dotted line) with the backscattered laser spectrum are presented.
}
\label{Fig9}
\end{figure}

In Figs.~\ref{Fig6}-\ref{Fig9}  we present our constraints on $W'$
couplings.  The SM
corresponds to
the origin and we vary the $W'$ 
couplings about it, showing the contours corresponding to a 95\% CL
deviation.  
Thus, $W'$ couplings lying within the limits would be 
indistinguishable from the SM while those outside would indicate 
statistically significant deviations from the SM.
 For simplicity we have taken $C_L^e=C_L^q$ and
$C_R^e=C_R^q$.  This assumption is satisfied for the SSM, the LRM and the 
KK model.
We indicate the 
couplings corresponding to those three models with a full star, a dot, and
an open star, respectively.
If one were making simultaneous measurements with different processes 
one could extract lepton couplings from one measurement and then use 
those as input to constrain the quark couplings with this process.
We also point out that the couplings are normalized differently here than
in 
Ref.\cite{wp1} . 
For comparison, the limits of our figures correspond to values  of 
approximately $L_f(W)$ and $R_f(W)\simeq 0.5$ in Figs. 12-17 of Ref. 
\cite{wp1} .

In Fig.~\ref{Fig6} we show the constraints on the couplings of a 750~GeV
$W'$ 
at a 500~GeV collider for different observables and
beam polarization. The results are given for the case of an integrated
luminosity of 500~fb$^{-1}$, including a 2\% systematic error.  One sees
that 
one can obtain interesting constraints even though the $W'$ mass is greater 
than the center of mass energy. The binned differential cross section, 
$d\sigma/d{p_T}_q$, gives the strongest constraint (solid line).  100\% 
right-handed polarization gives a strong 
constraint from $\sigma_R$ orthogonal to
that 
obtained from $d\sigma/d{p_T}_q$ (dashed line).  However, we see that
the 
constraints from $\sigma_R$ are seriously degraded for 90\% polarization
(dotted line).

Since we find that $d\sigma/d{p_T}_q$ gives the best limits we will
explore 
variations of machine parameters and $W'$ properties in
Figs.~\ref{Fig7}-\ref{Fig9}, using
that observable only. 

In Fig.~\ref{Fig7} we show the effect of different luminosities and of
including 
a systematic error. For the case of a 750~GeV $W'$ at a 500~GeV collider
illustrated in the figure, the SSM and KK models are distinguishable from
the standard model, even for the low luminosity case of 50~fb$^{-1}$, with
a
2\% systematic error included. On the other hand, the LRM is not
distinguishable even for the high luminosity case without any systematic
error included. This is consistent with the mass limits quoted for this
model in Table I, where $M_W' =$ 750~GeV is right at the limit of
discovery for the most favourable case.

In Fig.~\ref{Fig8} we show the constraints on the $W'$ couplings at a
500~GeV
collider for three
representative $W'$ masses of 0.75~TeV, 1~TeV, and 1.5~TeV. In
Fig.~\ref{Fig9}, for
the case of a 1.5~TeV $W'$, the constraints are shown for three different
collider energies of 0.5~TeV, 1.0~TeV, and 1.5~TeV. In both
Figs.~\ref{Fig8} 
and \ref{Fig9} we
have presented the case of an integrated luminosity of 500~fb$^{-1}$ with
a
2\% systematic error included. In each case shown in the figures, the SSM
and KK models are distinguishable from the standard model.

\section{Conclusions}

In this paper, we studied the sensitivity of the process 
$e\gamma\to\nu q$  to $W'$ bosons.  We used this process to 
find $W'$ mass
discovery limits and to see how well one could measure
the couplings of the $W'$  bosons expected in various
extensions of the standard model.  

For the discovery limits the highest reach was 
obtained by binning the $d\sigma/d{p_T}_q $ distribution.
For most models, 
the discovery reach of the backscattered laser process 
 is 
typically in the 2-10~TeV range depending on the 
center of mass energy, the 
integrated luminosity, and the assumptions regarding systematic errors.
These limits compare very favourably with other processes, including 
measurements at
the LHC. For the $e^+e^-$ process with WW photons, the reach is typically
in the 1-6~TeV range.

 For the $W_R$ boson of the LRM, for 
which LHC discovery limits are available, the discovery limits are
significantly lower.
For $g_R=g_L$, 
$M_{W'}$= 0.75, 1.2, and 1.6 TeV for $\sqrt{s}=$ 500, 1000, and 1500 
GeV respectively assuming $L_{\rm int}=500$~fb$^{-1}$ relative to a reach
of $~5.9$~TeV at the LHC.

Even for cases where  the discovery reach for $W'$'s 
with this process is not competitive with the reach of the LHC, precision 
measurements can give information on extra gauge boson couplings 
which complements that from the LHC.  In particular, if the LHC were to
discover 
a  $W'$ the process $e \gamma \to \nu q$ could be used to
constrain $W'$ couplings.  

We have demonstrated that this process has a great deal of potential in 
searching for the effects of extra gauge bosons.  As we stated at the 
outset the resolved photon approach was a useful approximation 
adequate for estimating the discovery potential.  However, when 
considering real data one must of course have exact calculations 
including radiative corrections.  With the knowledge that this 
process may be a good probe for new physics the motivation to perform 
more detailed calculations now exists.

\acknowledgments

This research was supported in part by the Natural Sciences and Engineering 
Research Council of Canada.
S.G. and P.K. thank Dean Karlen for useful discussions and Michael 
Peskin and Tom Rizzo whose conversations led to studying this process.
The work of M.A.D.\ was supported, in part, by the Commonwealth
College of The Pennsylvania State University under a Research
Development Grant (RDG).

\newpage

\newpage

%\newpage

%\newpage

%\newpage

%\newpage

%\newpage

%\newpage

%\newpage

\newpage
\begin{table}[t]
\caption{$W'$ 95\% C.L.\ discovery 
limits, in TeV, for the backscattered laser case obtained in the SSM, LRM
($\kappa=1$),
UUM, and the KK model using 
$d\sigma/dp_{T_{q}}$ as the observable. Results are presented 
for $\sqrt{s}=500$, 1000,  and 1500 GeV and for two luminosity 
scenarios, with and without a 2\% systematic error included.
}
\label{limitstabbl}
\vspace{0.4cm}
\begin{center}
\begin{tabular}{llllllllll}
%\hline
           &Lum.\ (fb$^{-1}$):& 50   & 500  & 50   & 500    \\
$\sqrt{s}$ &Sys. Err.: & 0\% 	& 0\% 	& 2\%  	& 2\%   \\
(GeV)      &Model   &  		&  	&   	&  	\\
\hline
500  & SSM    	    &2.3 &4.1 	& 1.9 	& 2.6   \\
     & LRM          &0.53 & 0.75 &0.51 & 0.63   \\
     & UUM          &2.3 & 4.1  &1.9 &2.6   \\
     & KK           &3.2 &5.7 & 2.7  & 3.6  \\
\hline
1000       &Lum.\ (fb$^{-1}$):& 200   & 500  & 200   & 500   \\
\hline
     & SSM   	    &4.6 &5.8  &3.7  &4.2   \\
     & LRM          &1.0 &1.2 &0.98 &1.1   \\
     & UUM          &4.6 &5.8 &3.7 &4.2  \\
     & KK           &6.6 &8.3 &5.2  &6.0   \\
\hline
1500       &Lum.\ (fb$^{-1}$):& 200   & 500  &200   & 500  \\
\hline
     & SSM	    &5.7  &7.2 &4.8 &5.6  \\
     & LRM          &1.4 &1.6  &1.3  &1.5   \\
     & UUM          &5.7  &7.2 &4.9  &5.7  \\
     & KK           &8.1 &10. &6.8  &8.0  \\
\end{tabular}
\end{center}
\end{table}

\newpage
\begin{table}[t]
\caption{$W'$ 95\% C.L.\ discovery 
limits, in TeV,
for the $e^+e^-$ case with WW photons obtained in the SSM, LRM
($\kappa=1$),
UUM, and the KK model using 
$d\sigma/dp_{T_{q}}$ as the observable. Results are presented 
for $\sqrt{s}=500$, 1000,  and 1500 GeV and for two luminosity 
scenarios, with and without a 2\% systematic error included.
}
\label{limitstabww}
\vspace{0.4cm}
\begin{center}
\begin{tabular}{llllllllll}
%\hline
           &Lum.\ (fb$^{-1}$):& 50   & 500  & 50   & 500    \\
$\sqrt{s}$ &Sys. Err.: & 0\% 	& 0\% 	& 2\%  	& 2\%   \\
(GeV)      &Model   &  		&  	&   	&  	\\
\hline
500  & SSM    	    &1.4 &2.5 	& 1.3 	& 1.9   \\
     & LRM          &0.38 & 0.54 &0.37 & 0.51   \\
     & UUM          &1.4 & 2.5  &1.3 &1.9   \\
     & KK           &2.0 &3.5 & 1.8  & 2.7  \\
\hline
1000       &Lum.\ (fb$^{-1}$):& 200   & 500  & 200   & 500   \\
\hline
     & SSM   	    &2.9  &3.6  &2.5  &3.0   \\
     & LRM          &0.74 &0.85 &0.72 &0.82   \\
     & UUM          &2.9 &3.6 &2.5 &3.0  \\
     & KK           &4.1 &5.1 &3.5  &4.2   \\
\hline
1500       &Lum.\ (fb$^{-1}$):& 200   & 500  &200   & 500  \\
\hline
     & SSM	    &3.6  &4.5 &3.2 &3.8  \\
     & LRM          &0.95 &1.1  &0.93  &1.1   \\
     & UUM          &3.6  &4.5 &3.2  &3.9  \\
     & KK           &5.1 &6.4 &4.5  &5.4  \\
\end{tabular}
\end{center}
\end{table}

\end{document}